\documentclass[doublecol]{epl2}

\title{Simulation of anyonic fractional statistics of Kitaev's toric model in circuit QED}
\shorttitle{Simulation of anyonic fractional statistics in cQED}

\author{Zheng-Yuan Xue}
\shortauthor{Z.-Y. Xue}

\institute{Laboratory of Quantum Information Technology, ICMP and
School of Physics and Telecommunication Engineering, South China
Normal University, Guangzhou 510006,  China}


\abstract{Since the anyonic excitations in the Kitaev toric model
are perfectly localized quasiparticles, it is possible to generate
dynamically the ground state and the excitations of the model
Hamiltonian to simulate the anyonic interferometry. We propose a
scheme in circuit QED to simulate the interferometry. The
qubit-cavity interaction can be engineered to realize effective
state control as well as the controlled dynamics of qubits, which
are sufficient to prepare the ground states, create and remove the
anyonic excitation, and simulate the anyonic interferometry. The
simplicity and high fidelity of the operations used open the very
promising possibility of simulating fractional statistics of anyons
in a macroscopic material in the near future.}

\begin{document}

\maketitle

\section{Introduction}
Anyons are exotic quasiparticles living in two dimensions with
fractional statistics \cite{anyon}.  A paradigmatic system for the
existence of anyons is a kind of so-called fractional quantum Hall
states \cite{fqhs}. Alternatively, artificial spin lattice models
are also promising for observing these exotic excitations
\cite{wenbook,kitaev,xue}. Recently, with the potential applications
in topological quantum computation, anyons have attracted strong
renewed interests
\cite{duan,zoller,youtopo,zhang,jiang,han,lucy,pachos,zhangj,cirac}.
However, a direct observation of fractional statistics associated
with anyon braiding is hard. For the Kitaev toric model, it has been
proposed \cite{han} an alternative, i.e., to generate dynamically
the ground state and the excitations of the model Hamiltonian,
instead of direct ground-state cooling, to simulate the anyonic
interferometry. The anyons are perfectly localized quasiparticles in
that model Hamiltonian, therefore one do not need a large system for
implementing their braiding operations. Indeed, small-scale system
for proof-of-principle simulation of the anyonic fractional
statistics was realized in optical systems
\cite{lucy,pachos,zhangj}. This kind of experiment represents an
important step toward the long pursued goal to simulate fractional
statistics of quasiparticles in a macroscopic material, which is
critical for future implementation of topological quantum computer,
providing the ability to scale up to large systems.

Superconducting circuit is one of the most promising candidates
serving as hardware implementation of quantum computers \cite{You}.
In this paper, we propose a scheme in circuit QED to simulate the
interferometry. The qubit-cavity interaction can be engineered to
realize effective state control as well as the controlled dynamics
of the  two-level systems, which are sufficient \cite{cirac} to
prepare the ground states of the toric Hamiltonian, create and
remove the anyonic excitation, and simulate the anyonic
interferometry.  The simplicity and high fidelity of the operations
used may open the very promising possibility of simulating
fractional statistics of anyons in a macroscopic superconducting
phase qubit in the near future.

\section{Quantum dynamics in circuit QED}

The superconducting charge qubit considered here consists of a small
superconducting box with excess Cooper-pair charges, formed by a
symmetric SQUID with the capacitance $C_{J}$ and Josephson coupling
energy $E_J$, pierced by an external magnetic flux $\Phi$. A control
gate voltage $V_g$ is connected to the system via a gate capacitor
$C_g$.  Focus on the charge regime, at temperatures much lower than
the charging energy [$E_{c}=2e^2/(C_g+2C_{J})$] and restricting the
induced charge [$\bar{n}=C_g V_g/(2e)$] to the range of $\bar{n}\in
[0,1]$, only a pair of adjacent charge states on the island are
relevant. The qubits are capacitively coupled to a transmission line
resonator which forms a 1D cavity. For simplicity, we here assume
that the cavity has only a single mode that plays a role. To obtain
maximum coupling strength, they are fabricated close to the voltage
antinodes of the cavity. This coupling is determined by the gate
voltage, which contains both the dc contribution and a quantum part.
The qubits are working at their optimal points, where they are
immune to the charge noise and possess long decoherence time. In the
qubit eigenbasis, neglecting fast oscillating terms using the
rotating-wave approximation, the Hamiltonian describes this scenario
now takes the usual Jaynes-Cummings form \cite{yale}
\begin{equation} \label{jc}
H_{JC} = \omega_r a^\dag a +{\nu \over 2}\sigma^z - g \left(a^\dag
\sigma^- +a\sigma^+  \right),
\end{equation}
where we have assume $\hbar=1$, $\nu$ is the qubit energy splitting,
$g$ is the coupling strength of a qubit to the cavity, $\omega_r$,
$a$ and $a^\dag$ is the frequency, annihilation and creation
operator of the cavity field, respectively. Note that the qubit
frequency can be tune within a large range by the external magnetic
field. Therefore, selected qubit-cavity interaction can be achieved.

The evolution operator of qubit-cavity interaction Hamiltonian in
Eq. (\ref{jc}) is given by
\begin{eqnarray}
\label{u} U(t)=\left(\begin{array}{cccc}
           1 & 0 & 0 & 0 \\
           0 & \cos\theta & -i\sin\theta & 0 \\
           0 & -i\sin\theta & \cos\theta & 0 \\
           0 & 0 & 0 & 1
         \end{array}
\right),
\end{eqnarray}
where $\theta=gt$, and it results in an oscillation between the
qubit and cavity states.

Meanwhile, driving in the form of
\begin{equation}\label{hd}
h=\varepsilon(t)a^\dagger e^{-i\omega_d t}+ \varepsilon^*(t) a
e^{i\omega_d t}
\end{equation}
on the resonator can be obtained \cite{yale} by capacitively
coupling it to a microwave source with frequency $\omega_d$ and
amplitude $\varepsilon(t)$. Depending on the frequency, phase, and
amplitude of the drive, different logical operations for qubit can
be realized.

To get fast gate, we work with large amplitude driving fields, where
quantum fluctuations are very small compare with the drive
amplitude, and thus the drive can be considered as a classical
field. In this case, it is convenient to displace the field
operators using the time-dependent displacement operator
\cite{scully}: $$D(\alpha) = \exp\left(\alpha a^\dagger - \alpha^* a
\right).$$ Choosing $i\dot\alpha = \omega_r\alpha +\varepsilon(t)
e^{-i\omega_d t}$ to eliminate the direct drive on the resonator, by
$h$, from the effective Hamiltonian, then the displaced Hamiltonian
for a qubit reads \cite{yale}
\begin{equation}\label{hd}
 H_D=\omega_r a^\dagger a +
{\nu   \over 2}\sigma^z  -  g \left[ \left(a+\alpha\right)\sigma^+ +
H.c.\right].
\end{equation}
 When the drive amplitude
is independent of time, and change to a frame rotating at
$\omega_d$, the displaced Hamiltonian reads
\begin{equation}\label{rotate}
H_\mathrm{RF} = \delta a^\dag a+ {\Delta\over 2} \sigma_z +
{\Omega\over 2} \sigma_x - g  \left(a \sigma^{+}  +a^\dag \sigma^{-}
\right).
\end{equation}
where $\delta=\omega_r-\omega_d$,  $\Delta=\nu-\omega_{dr}$, and
$\Omega=2g\varepsilon/\delta$ is the Rabi frequency.  In the
dispersive regime $\delta\gg g$,  the effective Hamiltonian is
\cite{yale}
\begin{equation}
\hat{H}_x=\delta\hat{a}^\dagger\hat{a}+\frac{\Delta+g^2/\delta}{2}\hat{\sigma}_z+\frac{\Omega}{2}\hat{\sigma}_x.
\label{x}
\end{equation}
By choosing $\Delta+g^2/\delta=0$, the Hamiltonian (\ref{x}) evolves
as a rotation around the $x$ axis. The gate speed scales as
$t_x\sim1/\Omega$, which have already been experimentally verified
\cite{rabi}.

We now turn to consider that the drive is sufficiently detuned from
the qubit $|\Delta|\gg\Omega$. Then, the effective Hamiltonian from
Eq. (\ref{rotate}) is \cite{yale}
\begin{equation}
\hat{H}_z=\delta\hat{a}^\dagger\hat{a}
+\frac{\chi}{2}\hat{\sigma}_z, \label{z}
\end{equation}
which generates rotations around $z$ axis at a rate
$\chi=\Delta+g^2/\delta+\Omega^2/(2\Delta)$, and the  gate speed
scales as $t_z\sim1/\chi$.

From the above analysis, we can see that switch on and off the
qubit-cavity coupling can be achieved by tuning the external driven
field. In this way, we can obtain individual manipulation of qubits.
It worth noticing that the dispersive regime in single-qubit
operation is induced by the detuning with respect to the driving,
which will not slow the gate speed as in the dispersive coupling of
qubits via virtual excitation of the cavity mode \cite{yale}.
Specifically, $t_x\sim1/g$ when $\epsilon\sim\delta/2$.

Note that the iSWAP gate between the cavity and $j$th qubit can be
obtained from evolution operator (\ref{u}) when $\tau=\pi/(2g)$.
Together with single-qubit rotations, it can be used to generate
cluster states more efficiently than that of controlled phase flip
gate \cite{tt,yu}. The operation used for the cavity and qubit
system is
\begin{eqnarray} \label{uc}
\label{uc} U^c&=&I \otimes Z_{(\pi/2)} \times U(\tau)\times I \otimes Z_{(\pi/2)} \nonumber\\
&=&\left(\begin{array}{cccc}
           1 & 0 & 0 & 0 \\
           0 & 0 & 1 & 0 \\
           0 & 1 & 0 & 0 \\
           0 & 0 & 0 & -1
         \end{array}
\right),
\end{eqnarray}
where $Z_{(\pi/2)}$ is a $\pi/2$ rotation around the $z$ axis of the
qubit state. This gate makes the two involved parties linked by a
controlled phase flip gate and at the same time it swaps their
states. It is worth noticing that both single-qubit and iSWAP
operation times are several $n$s which is much smaller compare to
typical decoherence time of the qubit \cite{alevel} and cavity
\cite{c}, both of them being on the order of $\mu$s.

\section{Ground states preparation}

\begin{figure}[tbp]
\includegraphics[width=8.5cm]{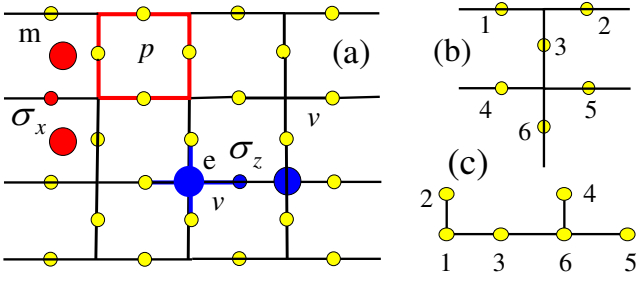}
\caption{(Color online) The toric model. (a). It is defined as the
ground state of a stabilizer Hamiltonian on a square lattice with
spins at the edges of a square lattice. The commuting vertices
(plaquettes), as indicated by a blue cross (red rectangle), are the
stabilizers. A single $\sigma_x$ ($\sigma_z$) gate on a spin,
indicated by a small red (blue) circle, can create a pair of
magnetic (electric) defects excitation, indicated by a big red
(blue) circle. (b). The smallest system for implementation of the
anyon braiding operation, the ground state of which is equivalent
under local single-bit operations to a graph state in (c).}
\label{toric}
\end{figure}

In the following, we show how to prepare the ground states of the
toric model \cite{kitaev} with this gate. The toric code is defined
as the ground level of a stabilizer Hamiltonian
\begin{equation}
H = - \sum_{v} A_{v} - \sum_{p} B_{p}
\end{equation}
on a square lattice with spins, realized as superconducting qubits
here, at the edges of a square lattice as shown in Fig.
(\ref{toric}a). The sum is over the mutually commuting stabilizers
$$A_{v} = \prod_{
  i \in v} \sigma^x_i, \quad B_{p} = \prod_{j \in p}
\sigma^z_j,$$ where $v$ runs over all vertices and $p$ over
plaquettes, as indicated by a blue cross and a red rectangle in Fig.
(\ref{toric}a). The ground state $|\varphi\rangle_g$ is
characterized by $A_{v}=B_{p}=1$. Plaquette and vertex excitations
$|\varphi\rangle_{e,m}$, also called magnetic and electric defects
and indicated in Fig. (\ref{toric}a) by bigger red and blue fulled
circles, are characterised by $B_{p} = -1$, $A_{v} = -1$ and  they
are mutual Abelian anyons. Therefore, apply single $\sigma_x$ or
$\sigma_z$ can create these quasiparticle excitations.  These
excitations always appear in pairs, at the ends of strings of
$\sigma_x$ and $\sigma_z$ operators applied on a ground state. To
apply $\sigma_x$ or $\sigma_z$ to a spin, we need to turn on the
qubit-cavity interaction in Eq. (\ref{x}) or (\ref{z}).

\begin{figure}[tbp]\centering
\includegraphics[width=6cm]{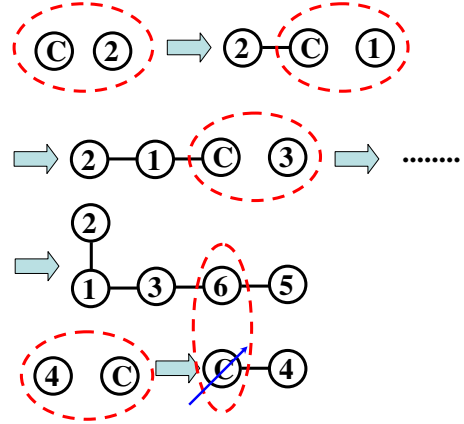}
\caption{(Color online) The process of generating the ground state
graph with six spins. Each dashed circle denotes a $U^c$ operation
and each bond means the two are linked by a controlled phase flip
gate. Circle with capital letter "C"  and numbers denote the cavity
and spins, respectively. The blue arrow across the circle represents
"X" measurement on the cavity.} \label{graph}
\end{figure}

For the toric model, the excitations are perfectly localized.
Therefore, for a proof-of-principle simulation of anyons and their
braiding statistics, the minimum implementation only needs six spins
\cite{han}, as the graph shown in Fig. \ref{toric}(b). The ground
state of which is equivalent under local single-bit operations to a
graph state in Fig. \ref{toric}(c). We next show how to prepare the
ground state. We work with a appropriate 3-body vertex and plaquette
operators along the boundary providing for a two-dimensional code
space. We start with a well defined state, i.e., the cavity and
spins 1, 2, 3, 5, and 6 are initialized to the state of
$|+\rangle=(|0\rangle+|1\rangle)/\sqrt{2}$ while spin 4 is prepared
in the ground state $|g\rangle$. The process is illustrated in Fig.
(\ref{graph}). We first sequentially apply $U^c$ on cavity and spins
2, 1, 3, 6, and 5, which results in the five qubit cluster state and
leave the cavity to the $|0\rangle$ state. As illustrated in the
Fig. (\ref{graph}), after each $U^c$, the cavity is at the right end
of the cluster. Then, apply $U^c$ on spin 4 to create an bipartite
entanglement. At last apply $U^c$ on spin 6 to fuse the two
entangled states followed by an $X$ measurement on the cavity state.
After these steps, the six spins are now prepared in the graph state
as in Fig. (\ref{toric}c). Measurement on the cavity can be
implemented by swapping the state of the cavity to another spin and
then measure the spin state.

\section{Controlled dynamics and anyon interference}
This operation in Eq. (\ref{uc}) can also be used to realize
controlled rotation $U_z=|0\rangle\langle0|\otimes\sigma_z+
|1\rangle\langle1|\otimes I$ of the qubit about the $z$ axis under
the control of the cavity states. It goes as following: 1) Prepare
the cavity state to the state of $(|0\rangle+|1\rangle)$. Here, we
need to engineer the cavity number states, which cab be achieved by
swapping an ancillary qubit  states with that of the cavity. 2)
Excite the qubit excited state $|e\rangle$ to an ancillary level
(other than $|g\rangle$ and $|e\rangle$). Ancillary levels have
already been used in a recent experimental demonstration of gates in
circuit QED \cite{alevel}. 3) Apply $U^c$ for the cavity and the
target spin for a time of $\pi/S$. Note that
$U_x=|0\rangle\langle0|\otimes\sigma_x+|1\rangle\langle1|\otimes I$
is equivalent to $U_z$ up to local single-qubit operation on the
spin \cite{jiang}: $U_x=HU_zH$. It is note that our controlled
operations are different from that of Refs. \cite{jiang,cirac}. The
essential of the controlled operations is to obtain state-dependent
dynamics, in this sense, the effect of our controlled operations are
the same as their's.

With these controlled operators and rotating gates of the qubits, it
is sufficient to create and remove the excitations and to simulate
the anyonic interferometry of the toric model.   To create a
superposition of the ground and excite state, we need an ancillary
degree of freedom, realized by the cavity here, then conditional
excitation
$\sim(|\varphi\rangle_g|1\rangle+\eta|\varphi\rangle_{e,m}|0\rangle)$
can be achieved, where $\eta$ is determined by the relative
amplitude of the controlled gate.  It is obvious that $\eta=1$ is
the controlled operation $U_{x,z}$. To apply $U_z$ to a spin, we can
get controlled magnetic defects excitation:
$U_x|\varphi\rangle_g|+\rangle\sim(|\varphi\rangle_g|1\rangle+|\varphi\rangle_m|0\rangle)$.
Similarly, one can also create controlled electric defects
excitation by
$U_z|\varphi\rangle_g|+\rangle\sim(|\varphi\rangle_g|1\rangle+|\varphi\rangle_e|0\rangle)$.

\begin{figure}[tbp]\centering
\includegraphics[width=8cm]{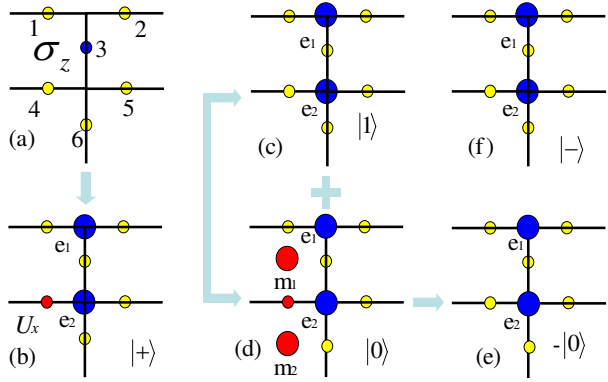}
\caption{(Color online) The minimal anyonic interferometry. The
label of spins (yellow small circles) is indicated in (a). Applying
a $\sigma_z$ gate to spin 3, creates two electric defects $e_1$ and
$e_2$, indicated as big blue circles in (b). With the cavity in
state $|+\rangle$, applying a $U_x$ gate on spin 4 conditionally
create two magnetic defects $m_1$ and $m_2$, i.e. the system now in
the superposition of (c) and (d). With the cavity in state
$|+\rangle$, applying $U_x$ gates on cavities and  spins 6, 5, 3 and
4 sequentially result in the superposition of (c) and (e), which is
equivalent to (f).} \label{inter}
\end{figure}

Once the toric code state is prepared, one can simulate the
fractional statistical phase of the anyons through a Ramsey-type
interference experiment. The simplest anyonic interferometry
\cite{han} is shown in Fig. (\ref{inter}). Apply a $\sigma_z$ gate
to spin 3, creates two electric defects $e_1$ and $e_2$. With the
cavity in state $|+\rangle$, apply a $U_x$ gate on spin 4 creates
the superposition states of (c) and (d). Apply $U_x$ gates on
cavities and spins 6, 5, 3 and 4 sequentially will move $m_2$ around
$e_2$ and finally fusing with $m_1$ to vacuum. By this braiding, the
-1 factor appears in the cavity state, i.e. the system now in the
superposition of (c) and (e), which is equivalent to (f). In this
interferometry the cavity state will change to
$|-\rangle=H|1\rangle$ from the state of $|+\rangle$. This phase
factor on the cavity state is solely due to the  mutually fractional
statistics of anyons and can be detected unambiguously in
experiment. This detection method is similar with that of Ref.
\cite{cirac}.

\section{Conclusion}

In summary, we propose a scheme in circuit QED to simulate the
fractional statistics of anyons of Kitaev's toric code model. The
qubit-cavity interaction can be engineered to realize effective
state control as well as the controlled dynamics of qubits, which
are sufficient to prepare the ground states, create and remove the
anyonic excitation, and simulate the anyonic interferometry. The
simplicity and high fidelity of the operations used here open the
very promising possibility of simulating fractional statistics of
anyons in a macroscopic material in the near future.

\acknowledgments I thank Prof. Shi-Liang Zhu for many helpful
suggestions. This work was supported by the NSFC (No. 11004065), the
NSF of Guangdong Province (No. 10451063101006312), and the Startup
Foundation  of SCNU (No. S53005).

\end{document}